\documentstyle[12pt,epsf]{article}

\begin{document}

\baselineskip=18pt plus 0.2pt minus 0.1pt

\makeatletter

\@addtoreset{equation}{section}
\renewcommand{\theequation}{\thesection.\arabic{equation}}
\renewcommand{\thefootnote}{\fnsymbol{footnote}}
\newcommand{\nn}{\nonumber}
\newcommand{\del}{\partial}
\newcommand{\vs}[1]{\vspace*{#1}}
\newcommand{\hs}[1]{\hspace*{#1}}
\newcommand{\wt}[1]{\widetilde{#1}}
\newcommand{\bm}[1]{\mbox{\boldmath $#1$}}
\newcommand{\bpi}{{\bf \Pi}}
\newcommand{\E}{{\bf E}}
\newcommand{\B}{{\bf B}}
\newcommand{\z}{\hat{\bf x}_3}
\newcommand{\r}{\hat{\bf r}}
\newcommand{\g}{g_{\rm st}}

\makeatother

\begin{titlepage}
\title{
\hfill\parbox{4cm}
{\normalsize KUNS-1576\\{\tt hep-th/9905162}}\\
\vspace{1cm}
Born-Infeld Dynamics in Uniform Electric Field\\
}
\author{
Koji {\sc Hashimoto}\thanks{
Supported in part by a Grant-in-Aid for Scientific
Research from the Ministry of Education, Science and Culture of Japan 
(\#3160). E-mail address: {\tt hasshan@gauge.scphys.kyoto-u.ac.jp}}
\\[40pt]
{\it Department of Physics, Kyoto University, Kyoto 606-8502, Japan}
}
\date{\normalsize May, 1999}
\maketitle
\thispagestyle{empty}

\begin{abstract}
\normalsize
We investigate various properties of classical configurations of
the Born-Infeld theory in a uniform electric field. This system
is involved with dynamics of (F,D$p$) bound states, which are bound
states of fundamental strings and D$p$-branes. The uniform electric
field can be treated as a constraint on the asymptotic behavior of the
fields on the brane. BPS configurations in this theory correspond to
fundamental strings attached to the (F,D$p$) bound state, and are
found to be stable due to force balance. Fluctuations around these
stable configurations are subject to appropriate Dirichlet and Neumann 
boundary conditions which are identical with the ones deduced from the
ordinary worldsheet description of the attached string. Additionally,
non-BPS solutions are studied and related physics are discussed.

\end{abstract}
\end{titlepage}

\section{Introduction}
\label{sec:intro}

Recent developments in string theory are greatly indebted to the
discovery of the importance of D-branes \cite{Pol}. Many kinds of
brane configurations have been studied, partly owing to the fact that
various dimensional gauge theories live on the D-branes. The
low-energy effective field theory on the D-brane is the Born-Infeld
theory \cite{BI, CLNY, BIeff, L}, a non-linear electrodynamics. This
particular non-linear system exhibits interesting properties,
and these can be well understood in terms of string theory. One of the
noteworthy facts on this system is that a deformed part of the surface
of the brane (parameterized by scalar fields on the brane)
corresponds to other type of branes, {\it i.e.}, intersecting branes
can be described by gauge theories \cite{CM,Gibb}. Using this
correspondence, new soliton solutions in gauge theories
have been found \cite{HHS} motivated by particular brane
configurations, and on the other hand, non-perturbative quantities in
string theory have been studied \cite{CM, death, death2} using
the Born-Infeld theories.

In this paper, we concentrate on the basic properties of the 
worldvolume Born-Infeld system in a uniform electric 
field\footnote{Related issues are found in Refs.\ \cite{add}.}. 
A
D$p$-brane with the uniform electric field can be identified with a
bound state of the D$p$-brane and fundamental strings (called an
(F,D$p$) bound state) \cite{Arf, Jab, Lu1}. We consider a fundamental
string ending on this bound state, as a generalization of
Refs. \cite{CM, Gibb}. Though the supergravity solution representing
this (F,D$p$) bound state has been constructed recently \cite{Lu2},
one of the missing important ingredients is explicit supergravity
solutions representing intersecting branes \cite{inter}. In the light
of the understanding of the AdS/CFT correspondence, the analysis in
the gauge theory side is of importance.

The organization of this paper is as follows.
In Sec.\ \ref{sec:stable}, we study the BPS property and stability of 
the configuration. Then in Sec.\ \ref{sec:bc}, we investigate the
(Dirichlet and Neumann) boundary conditions for the attached string by
means of Born-Infeld equations of motion, and see that these boundary
conditions are consistent with the ones deduced from the viewpoint of
string worldsheet \`{a} la Polchinski \cite{DLP}. Motivated by the
fact that a ``throat'' type of non-BPS solutions in this Born-Infeld
theory is involved with the decay of brane - anti-brane system
\cite{CM, death}, we analyze non-BPS configurations in the uniform
electric field in Sec.\ \ref{sec:anni}. The final section is devoted
to discussions, in paticular on the properties of these non-BPS
configurations. 


\section{Stability of BPS configuration}
\label{sec:stable}

\begin{figure}[htdp]
\begin{center}
\begin{minipage}{80mm}
\begin{center}
\leavevmode
\epsfxsize=60mm
\epsfbox{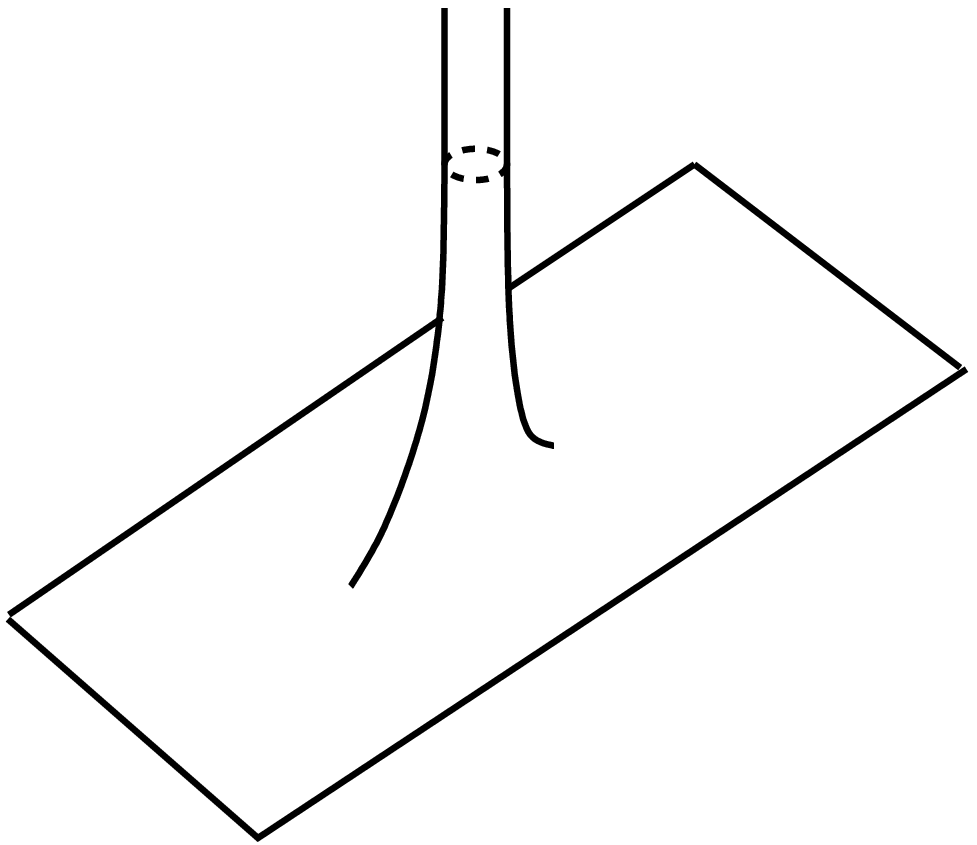}
\caption{``Tube-like'' configuration of D$p$-brane surface
  representing a string. Now the D$p$-brane is tilted, in order to be 
  a BPS configuration, under the uniform electric field.}
\label{fig:tube}
\end{center}
\end{minipage}
\hspace{5mm}
\begin{minipage}{70mm}
\begin{center}
\leavevmode
\epsfxsize=70mm
\put(-10,160){$X_9$}
\put(190,60){$x_3$}
\put(120,40){$f_{\rm string}$}
\put(70,15){$f_{\rm source}$}
\epsfbox{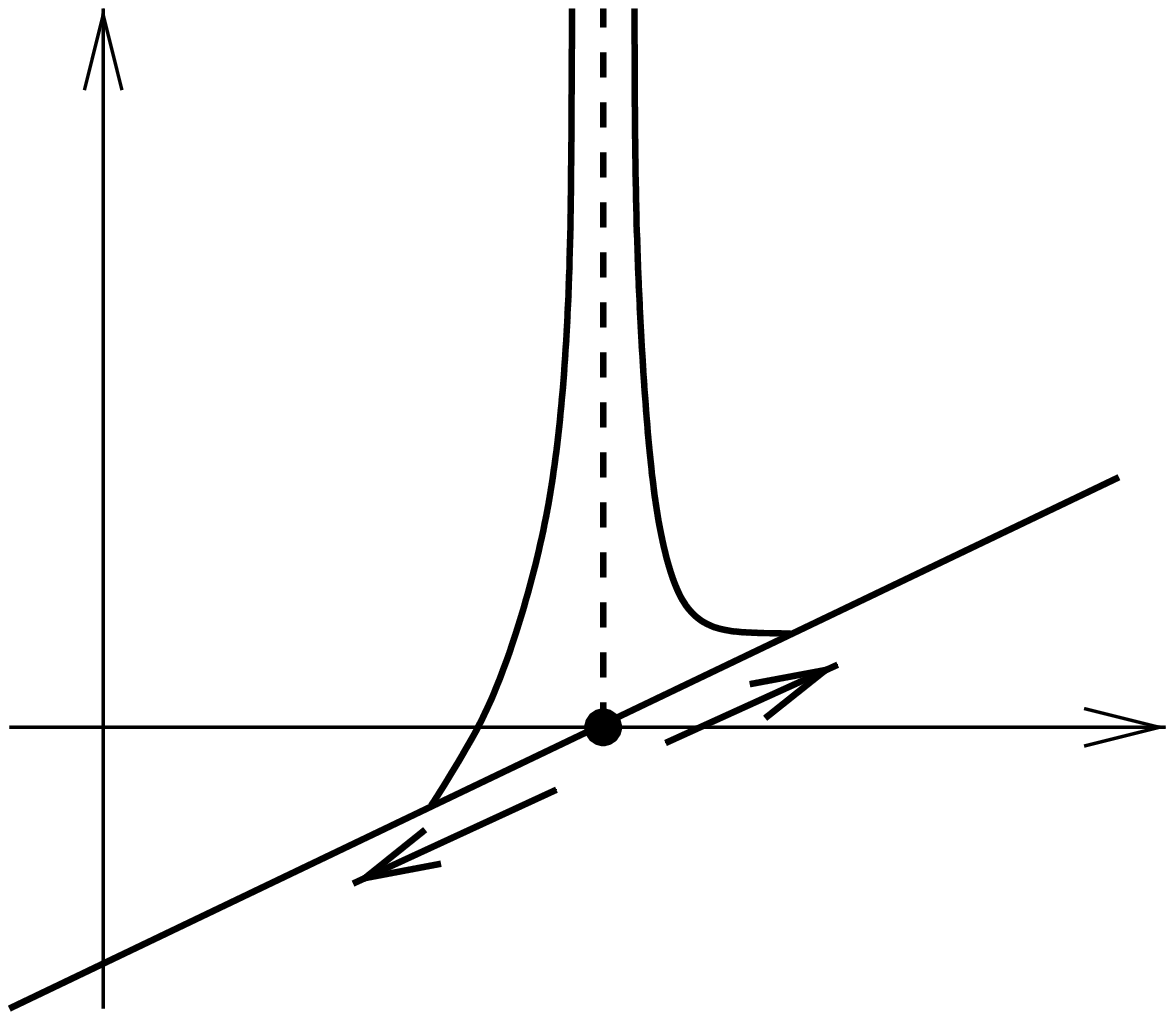}
\caption{Force balance of the configuration. }
\label{fig:bal}
\end{center}
\end{minipage}
\end{center}
\end{figure}

First, let us see how the uniform electric field and a point charge in
it are allowed as stable BPS configurations in the low-energy
effective theory of a D$p$-brane extending along the directions
$(012\cdots p)$. The argument on the linearized version of this system
given in Refs.\ \cite{CM} and \cite{Gibb} tells us that half of the
worldvolume supersymmetries are preserved when the fields on the brane
satisfy the following BPS condition 
\begin{eqnarray}
\label{susy}
  F_{0\mu}=\alpha F_{9\mu}, 
 \quad \mbox{with} \quad \alpha=\pm 1,
\end{eqnarray}
where $F_{9\mu}$ should be understood with a scalar field $A_9\equiv
X_9$, and the index $\mu$ runs $0, 1, \cdots, p$, directions parallel
to the brane. This BPS equation (\ref{susy}) is expected to be
derived also from the non-linear Born-Infeld theory \cite{susyBI}.
Under the relation (\ref{susy}), equations of motion of the
Born-Infeld theory agree with the ordinary ones in the Maxwell-scalar
system. As a solution\footnote{We choose $\alpha=-1$ and $p\geq 3$ in
  this paper.}, it is possible to generalize the solution adopted in
Ref.\ \cite{CM} so as to include trivially a uniform electric
field\footnote{This defines an exact conformal field theory, since the
  derivation in Ref.\ \cite{Th} depends not on the explicit form of
  the scalar potential, but only on the BPS relation (\ref{susy}).}: 
\begin{eqnarray}
\label{conf}
  X_9=-A_0= -\frac{c_p}{r^{p-2}} + Ex_3.
\end{eqnarray}
This configuration represents a charged particle in the background
electric field $\E=E\hat{\bf x}_3$ which is uniform on the
D$p$-brane\footnote{A point electric charge in a uniform 
  background magnetic field is not a BPS configuration. A
  BPS configuration with a magnetic point source in a uniform magnetic
  field, which corresponds to a D-string ending on  the D3-brane, will
  be discussed in Sec.\ \ref{sec:dis}. An issue concerning the uniform 
  magnetic field is found in Ref. \cite{KYL}.}.
As seen from Eq.\ (\ref{conf}), the attached fundamental string is not 
perpendicular to the D$p$-brane, because of the uniform field strength
(see Fig.\ \ref{fig:tube}). In other words, the D$p$-brane is now
tilted in order to preserve some supersymmetries.

From the viewpoint of target space supersymmetries, it is also
possible to see that the configuration given by Eq.\ (\ref{conf})
preserves a part of the supersymmetries to be stable. For simplicity,
consider the case of $p=3$. Taking T-dualities in two directions
along $x_1$ and $x_2$, then the tilted D-string (on which a constant
field strength exists) appears. This D-string can be interpreted as a
dyonic string carrying both NeveuSchwarz-NeveuSchwarz (NS-NS) and
Ramond-Ramond 2-form charges \cite{Wit}. Following the argument
developed in Ref.\ \cite{SENNET}, at the spatial infinity the
conserved supercharges are 1/4 of the original ones, under the
existence of this dyonic string and a fundamental string perpendicular
to the $(123)$-plane. The T-duality transformation does not change the
number of preserved supersymmetries, hence the configuration of Eq.\
(\ref{conf}) preserves 1/4 of the original target space
supersymmetries (at least at the spatial infinity). 

Although this configuration is stable because of its BPS nature,
naively one might be afraid of instability due to an electric force on
the point charge in the uniform electric field.  Actually, the charge
at $r=0$ is $(p\!-\!2)c_p\Omega_{p-1}$, and the force along the tilted 
direction can be read as  
\begin{eqnarray}
\label{fsou}
f_{\rm source} = -
(p\!-\!2)c_p \Omega_{p-1}\cdot T_p E \cdot \frac{1}{\sqrt{1+E^2}}.
\end{eqnarray}
Note that the strength of the ``physical'' electric field\footnote{The 
  charge quantization is calculated using the action 
  $$ T_p\int \! d^{p+1}\sigma 
   \sqrt{-\det (h_{\alpha\beta}+F_{\alpha\beta})}, $$
  therefore the Gauss law derived from this action is $\nabla\cdot 
  (T_p\E)=0$.} 
is $T_p E$, where $T_p$ is the tension of the D$p$-brane. The final
factor in Eq.\ (\ref{fsou}) is for extracting a  
component of the force along the tilted direction. Now, what happens
is that this force (\ref{fsou}) is exactly cancelled by 
$T_{\rm f}=1/2\pi$, the tension of the fundamental string emanated
from the D$p$-brane (see Fig.\ \ref{fig:bal}). Actually, the
contribution of the string tension along the tilted direction is
\begin{eqnarray}
f_{\rm string}=\frac{E}{\sqrt{1+E^2}} \frac{1}{2\pi},
\end{eqnarray} 
and using the charge quantization condition
$(p\!-\!2)c_p\Omega_{p-1}=g_{\rm st} 
(2\pi)^{p-1}$ (see Ref.\ \cite{CM}) and $T_p=1/(2\pi)^pg_{\rm   st}$, 
it is easy to see the force balance as
\begin{eqnarray}
  f_{\rm source} + f_{\rm string} =0.
\end{eqnarray}
This is consistent with the stability expected from the BPS nature of
the configuration.

\section{Boundary conditions}
\label{sec:bc}

\subsection{Worldsheet picture}
\label{sec:WS}

One of the remarkable properties of the description of the
intersecting strings in the Born-Infeld system is that the boundary
conditions for the 
attached strings can be reproduced \cite{CM, LPT, Rey}. Now, our
interest is not the pure D-branes but the (F,D$p$) bound state. In
this case, the boundary condition in the usual worldsheet picture are
known to be changed. We shall study this first, and check the
consistency with the Born-Infeld theory later.

In the worldsheet approach 
of D-branes by Polchinski et al.\ \cite{DLP}, the boundary conditions
of a fundamental string attached to the D-brane on which a uniform
gauge field strength exists are \cite{CLNY, L, Arf}
\begin{eqnarray}
\left(
\del_\sigma X^\mu - F^\mu_{\;\; \nu}\del_\tau X^\nu
\right)
\Biggm|_{\sigma=0}=0\quad
\mbox{($\mu=0,\cdots, p$)},
\label{mixed}\\
\del_\tau X^i\Biggm|_{\sigma=0}=0\quad
\mbox{($i=p\!+\!1,\cdots, 9$)}.
\end{eqnarray}
For the directions transverse to the D$p$-brane (Dirichlet
directions), the Dirichlet boundary conditions do not change in spite
of the existence of the electric field on the D$p$-brane. On the other
hand, for the Neumann directions, the uniform electric field
$F_{03}\equiv E$ yields so-called ``mixed'' boundary conditions, which
relate values of two or more scalars at the boundary. How can these
mixed boundary conditions be understood in terms of only a single
particular scalar, say $X^3$ ? Defining new coordinate scalars as 
$X^{\pm}\equiv \left(X^0\pm X^3\right)/\sqrt{2}$, then the above
boundary conditions (\ref{mixed}) along the plane spanned by $x^0$ and
$x^3$ reads \cite{Bachas} 
\begin{eqnarray}
  \left(\del_\sigma X^\pm \pm E\del_\tau X^\pm\right)
\Biggm|_{\sigma=0}=0.
\end{eqnarray}
A wave solution which respects this phase shift at the boundary
$\sigma=0$ is easily found as 
\begin{eqnarray}
   X^\pm=A_\pm
   \Bigl(
     \exp\left[i(\tau+\sigma\mp\alpha/2)\right] + 
     \exp\left[i(\tau-\sigma\pm\alpha/2)\right]
   \Bigr)
\end{eqnarray}
where the phase shift $\alpha$ is defined by a relation
$E=\tan(\alpha/2)$, and $A_\pm$ is the amplitude of the wave,
normalized as $|A_\pm|=1$. Hence the expression for the scalar $X^3$
is 
\begin{eqnarray}
  X^3=\sqrt{2}  \left( X^+ - X^- \right)=
\sqrt{2}\cos(\alpha/2)
  \Bigl( \exp[i(\tau+\sigma)] + \exp[i(\tau-\sigma)]  \Bigr).
\label{boux3}
\end{eqnarray}
Here we have chosen $A_+ = -A_- =1$, since this choice satisfies a
condition that the amplitude of the wave coming in is equal to the one 
of the wave going out. The expression (\ref{boux3}) indicates that the 
boundary condition for $X^3$ is purely Neumann-type, not the mixed
type. In the following, we will see that the above boundary conditions 
are reproduced in the Born-Infeld analysis.

\subsection{Transverse (Dirichlet) directions}
\label{sec:tra}

First, let us investigate the Dirichlet directions. For simplicity, we 
shall concentrate on the case $p=3$ hereafter. Denoting the
fluctuation in a direction transverse to both the attached string
and the D3-brane as $\eta$, the equation of motion for this
fluctuation is given by \cite{CM} 
\begin{eqnarray}
\label{callan}
  \left(1+|\nabla X|^2\right)\ddot{\eta}-\Delta \eta =0.
\end{eqnarray}
Substituting the configuration under consideration (\ref{conf}) and
putting the time dependence of the fluctuation $\exp (-i\omega t)$,
then Eq.\ (\ref{callan}) becomes
\begin{eqnarray}
  \left[\left(
      1+E^2 + \frac{2\pi E\g}{r^2}\cos \theta + \frac{\pi^2 \g^2}{r^4} 
          \right)\omega^2 +\Delta  \right]\eta=0,
\label{delta}
\end{eqnarray}
where $r\equiv\sqrt{x_1^2+x_2^2+x_3^2}$ and
$\cos\theta\equiv x_3/r$. Now one can see that the dependence on
$\theta$ appears in the equation, therefore the solution of this
equation cannot be spherically symmetric. We should take into account
the $\theta$-dependence of the solution.  

Assuming that the solution does not depend on another variable
$\varphi$ in polar coordinates, we can expand $\eta$ by Legendre
functions, a system of orthogonal functions, as 
\begin{eqnarray}
  \eta(r,\theta) = \sum_{l=0}^\infty \eta_l(r)P_l(\cos \theta).
\label{exp}
\end{eqnarray}
We require that if one takes the limit $E\rightarrow 0$ then the
solution should recover the one given in Ref.\ \cite{CM}. It is
possible to construct such a solution, and the result is (see
Appendix)
\begin{eqnarray}
\label{sol}
  \eta(r,\theta) = \hs{-10pt}&&\eta_0^{(0)}(r)\cdot
  \left(
    1+\sum_{l=1}^\infty P_l(\cos\theta)
    \left[
      (2\pi E\g\omega^2)^l \prod_{i=1}^l\frac{1}{(i+1)(2i-1)}
    \right]
  \right)\nn\\
&&+(\mbox{higher order terms}).
\end{eqnarray}
In the RHS, the ``higher order terms'' consist of terms which do 
not contribute to the phase of the total wave flux (the magnitude of
those higher order waves dump as taking the limit $r\rightarrow
0$. The region $r\sim 0$ corresponds to the tip of the tube in Fig.\
\ref{fig:tube}, where the initial and final states of the wave are
defined on the fundamental string.) The spherically symmetric factor
$\eta_0^{(0)}(r)$ is the solution of the equation 
\begin{eqnarray}
  \left[\left(1+\frac{\kappa^2}{y^4}\right)+\frac{d^2}{d y^2}
  \right]\eta_0^{(0)}(r)=0,
\label{kappa}
\end{eqnarray}
where we have put $\kappa^2\equiv (1+E^2)\pi^2 \g^2 \omega^4$ and
$y\equiv \pi \g\omega/r$. This is exactly the same one obtained in
Ref.\ \cite{CM}, except for the $E$-dependence of $\kappa$. Using the
tortoise-like coordinate 
\begin{eqnarray}
\xi(y)\equiv \int_{\sqrt{\kappa}}^y \sqrt{1+\kappa^2/y^4},
\end{eqnarray}
Eq.\ (\ref{kappa}) can be rewritten as a Schr\"odinger-type equation
\begin{eqnarray}
  \left(
    -\frac{d^2}{d\xi^2} + V(\xi)
  \right)\wt{\eta}=\wt{\eta}, \quad \mbox{where}\quad 
\wt{\eta}\equiv (1+\kappa^2/y^4)^{1/4}\eta.
\end{eqnarray}
The potential $V(\xi)$ approaches to a delta-function with infinite
area as one goes to the weak coupling limit $\g\rightarrow 0$. Thus
the solution (\ref{sol}) is subject to the Dirichlet boundary
condition at the weak coupling limit, as expected by the worldsheet
prescription in Sec.\ \ref{sec:WS}. 

\setcounter{footnote}{0}
\subsection{Longitudinal (Neumann) directions}
\label{sec:long}

For the string fluctuation in the longitudinal directions, which is
described by the fluctuation of the scalar $X^9$, we shall follow the
argument given in Ref.\ \cite{savisavi}. Turning on fluctuations of
both the gauge field and the scalar field $X^9$, the authors of Ref.\ 
\cite{savisavi} obtained the same equation (\ref{callan}), for the 
fluctuation of the gauge field, $\eta=\delta A_i$. The fluctuation of
the scalar field, $\delta X^9$, is related to $\delta A_i$ as 
\begin{eqnarray}
\label{gauge}
\del_i A_i + \del_t {\delta X^9}=0.
\end{eqnarray}
Therefore the boundary conditions for the $\delta X^9$ was found to be
Neumann-type. 

As discussed in Ref.\ \cite{savisavi}, among the various modes in
$\delta X^9$, the physical mode which precisely corresponds to the
fluctuation along the D-brane is the one carrying the angular momentum
$l=1$ in the worldvolume language. The solution (\ref{sol}) is
composed of the modes of all $l$, 
hence it may seem to be difficult to extract only the physical mode
mentioned above. 
In order to get the physical fluctuation along $x^3$, we shall turn on 
only $\delta A_z$. Then the relation (\ref{gauge}) gives 
\begin{eqnarray}
-i\omega\delta X^9 
\left(=\frac{\del}{\del z}A_z \right)
= \frac{z}r
\left(
  \frac{\del}{\del r}\eta_0^{(0)}(r)
  +\frac23 (\pi Eg\omega^2)^2\eta_0^{(0)}(r)
\right)+\cdots.
\label{neum}
\end{eqnarray}
The second term in the parenthesis in the RHS stems from
the $l=2$ excited part in the solution (\ref{sol}), and this mode
actually has Dirichlet property. However, taking the weak coupling
limit, this term vanishes (and this is also the case for other terms
denoted by ``$\cdots$'' in Eq.\ (\ref{neum}), which indicate
unphysical $l\neq 1$ modes). 

The other longitudinal directions (along $x^1$ and $x^2$) can be
analyzed in the same way. Summing up all together, we conclude that
in the weak coupling limit, for the fluctuations along the
longitudinal directions the boundary conditions are Neumann-type, as 
expected from the worldsheet picture.

Due to the change of the frequency parameter $\kappa$ in the
differential equation, there remains $E$-dependence at finite coupling
$\g$. Calculating the transmitting amplitude, the total power 
emanated from the end of the attached string \cite{savisavi} is now
$(1\!+\!E^2)$ times the ordinary flux $\omega^4 g^2$. This result is
natural in a sense that in the Born-Infeld system the speed of light
changes under the uniform field strength background. If we turn on the 
background adopted in this paper in the Born-Infeld-scalar system, the
velocity of the fluctuation becomes $1/\sqrt{1\!+\!E^2}$, as seen when
we neglect the terms originated from the point source in the
differential equation (\ref{delta}). This change can be viewed also 
as a change of the frequency $\omega$, therefore, this results in the
change of the total energy flux.

\vspace{10mm}
\section{Analysis for non-BPS solutions}
\label{sec:anni}

Particular non-BPS solutions of the non-linear Born-Infeld system are
fascinating. The first reason is that, they are analogue of the first
proposal by Born and Infeld \cite{BI} (called ``BIon''or ``pinched''
solution). The second one is that, they are concerned with the brane -
anti-brane annihilation \cite{CM, death} (called ``throat'' solution,
or ``(charged) catenoidal'' solution in Ref.\ \cite{Gibb}). Now our 
interest is mainly on the second respect, for the case of the 
(F,D$3$) bound state\footnote{The brane - anti-brane annihilation can
  be described also by tachyon condensation \cite{tac}. A recent
  result on the non-perturbative tachyon potential \cite{Hyaku} is
  based on the Matrix theory, in which generally there
  exists a uniform gauge field strength on the constructed brane.}.

We treat only static configurations. Differential equations which the
scalar $X$ ($\equiv X_9$) and the electric field $\E$ satisfy are as
follows \cite{CM}: 
\begin{eqnarray}
&&\nabla\cdot\bpi=0,
\label{divp}\\
&&\nabla\cdot\left(
\frac{\nabla X +\bpi(\bpi\cdot\nabla X)}
{\sqrt{1+|\nabla X|^2 + |\bpi|^2 + (\bpi\cdot\nabla X)^2}}
\right)=0.
\label{lap}
\end{eqnarray}
Here $\bpi$ is the canonical momentum associated with the gauge
field $A_{1,2,3}$, defined by\footnote{
Substituting the explicit representation of $\bpi$ (Eq.\ (\ref{pi}))
into Eq.\ (\ref{lap}), we obtain 
\begin{eqnarray}
\label{anot}
\nabla\cdot\left(
\frac{\nabla X (1-|\E|^2) +\E(\E\cdot\nabla X)}
{\sqrt{(1-|\E|^2)(1+|\nabla X|^2) + (\E\cdot\nabla X)^2}}
\right)=0.
\end{eqnarray}
Combined with Eq.\ (\ref{divp}), this equation (\ref{anot}) is 
invariant under the boost transformation in the plane spanned by
$\phi$ and $X$ \cite{Gibb}, where $\phi$ is the electro-static
potential, $\nabla \phi\equiv \E$. The throat solution in Ref.\
\cite{CM} can be easily constructed using this invariance. Let us
pursue this way of construction in our case. First, Put
$\E=0$ in Eqs.\ (\ref{divp}) and (\ref{anot}). Then the equation which 
should be solved is only one:
\begin{eqnarray}
\label{anot2}
\frac{\nabla X}{\sqrt{1+|\nabla X|^2 }}
=a\z + \frac{b}{r^2}\r,
\end{eqnarray}
since Eq.\ (\ref{divp}) is trivial due to the vanishing of the
momentum $\bpi$. Substituting $a=0$ leads to the throat solution in 
Ref. \cite{CM}.
Although this method to find a solution is a clever one, we cannot use
it for our purpose.  This is because the solution of our interest
should satisfy the boundary condition (\ref{bou}), and to attain this
boundary behavior one must perform the boost on the solution 
(\ref{anot2}) with infinite velocity. Then the boosted solution
diverges and becomes meaningless.

One of the other ways to solve Eqs.\ (\ref{divp}) and (\ref{anot}) is
to put $\E=\z$ (and hence $|\E|^2=1$). This assumption makes these
equations easy a great deal, and the solution is 
\begin{eqnarray}
\label{solc}
  \nabla X=
\left({c+\frac{A\cos\theta}{r^2}}\right)^{-1} \!
\left({1+\frac{cA\cos\theta}{r^2}+\frac{q^2}{r^4}}\right)
\z + \frac{-A}{r^2}\r
\end{eqnarray}
where $c$ is an integration constant. When $c\neq 1$, even with the
above boost, this solution can not be brought into the form satisfying 
the boundary condition (\ref{bou}). Therefore, we must put the
value of $c$ equal to 1. We find that this is precisely the special
case of the general solution expressed by Eq.\ (\ref{sol!}). Putting
$B$ equal to zero in Eq.\ (\ref{sol!}), we reproduce the solution 
(\ref{solc}).
} 
\begin{eqnarray}
\label{pi}
\bpi=\frac{\E(1+|\nabla X|^2) -\nabla X(\E\cdot\nabla X)}
{\sqrt{(1-|\E|^2)(1+|\nabla X|^2) + (\E\cdot\nabla X)^2}}.
\end{eqnarray}
In the BPS limit $\E=\nabla X$, these two equations (\ref{divp}) and
(\ref{lap}) reduce to the ordinary Gauss law, $\nabla\cdot\E=0$. 

Since we want to deal with the separated
parallel brane - anti-brane system,  we shall fix the boundary
of the brane at the spatial infinity by identifying them with the BPS 
case :
\begin{eqnarray}
\label{bou}
\E\sim \nabla X \sim E\z \qquad {\rm at } \qquad r\sim \infty.  
\end{eqnarray}
Substituting this boundary condition into Eq.\ (\ref{pi}), then we
have $\bpi \sim E\z$ at the spatial infinity of the worldvolume. The
same argument is applied for the inside of the large parenthesis of
Eq. (\ref{lap}), therefore the solution of the Eqs.\ (\ref{divp})
and (\ref{lap}) respecting the boundary conditions are written as
follows:
\begin{eqnarray}
\bpi = E\z + \frac{A}{r^2}\r \\
\frac{\nabla X +\bpi(\bpi\cdot\nabla X)}
{\sqrt{1+|\nabla X|^2 + |\bpi|^2 + (\bpi\cdot\nabla X)^2}}
= E\z + \frac{B}{r^2}\r 
\end{eqnarray}
For simplicity, we put both $A$ and $B$ positive.
After some manipulations we find the general solution
\begin{eqnarray}
\label{sol!}
  \nabla X=\frac{P\z + Q\r}{\sqrt{R}},
\end{eqnarray}
where three functions $P$, $Q$ and $R$ are assumed to depend only on
$r$ and $\theta$ in a polar coordinate system ($\cos\theta=x_3/r$),
and their explicit forms are:
\begin{eqnarray}
&& P(r,\theta)\equiv E\left(1+\frac{(A\!-\!B)E\cos\theta}{r^2} 
             +\frac{A(A\!-\!B)}{r^4}\right),\\ 
&& Q(r,\theta)\equiv\frac1{r^2}
       \left[\left\{(1\!+\!E^2)B - E^2A\right\}
               -\frac{EA(A\!-\!B)\cos\theta}{r^2}\right],\\
&& R(r,\theta)\equiv 1 + \frac{2E(A\!-\!B) \cos\theta}{r^2} \nn\\
&&\hs{100pt} + \frac{(A\!-\!B)\left\{(1\!-\!E^2) A + (1\!+\!E^2) B 
               + E^2 (A\!-\!B) \cos^2\theta\right\}}{r^4}.
\hs{20pt}{}
\end{eqnarray}
As is easily checked, at the spatial infinity $r\rightarrow \infty$,
these three functions have asymptotic behavior $R\rightarrow 1$,
$Q\rightarrow 0$, $P\rightarrow E$, and hence the boundary condition 
(\ref{bou}) is satisfied.

As expected from the case with no uniform electric field 
\cite{CM, death}, the BPS limit corresponds to the relation
$A=B$. Actually, in this limit, three functions are simplified as
$P=E$, $Q=A/r^2$ and $R=1$, then we reproduce the BPS solution
(\ref{conf}).

We can characterize the configuration of the brane by the location
where $\nabla X$ diverges. This is defined by the root of the
denominator of the solution, $\sqrt{R}$,
\begin{eqnarray}
  r^4 R= 
  \left(
    r^2-E(B\!-\!A)\cos\theta
  \right)^2
 -(B\!-\!A)\left\{(A\!+\!B)+E^2(B\!-\!A)\right\}=0.
\end{eqnarray}
If $B>A$, there is a solution 
\begin{eqnarray}
  (r_{\rm critical})^2 =
  \sqrt{(B\!-\!A)\left\{(A\!+\!B)+E^2(B\!-\!A)\right\}}+
  E(B\!-\!A)\cos\theta.
\label{cri}
\end{eqnarray}
At this radius $r_{\rm critical}$, the gradient $\nabla X$
diverges and the brane forms a ``throat''. Inside the critical surface
defined by Eq.\ (\ref{cri}), $r<r_{\rm critical}(\theta)$, the
solution is not defined. Thus the brane has a hole. This critical
surface has a shape of stretched sphere, due to   
the existence of the uniform electric field. On the 
other hand, if $A>B$, the solution is a generalization of 
``BIon''\footnote{In the case $E^2>1$, there exists
  the ``fourth'' phase, in addition to the above three situations
  (throat, BPS and BIon). This phase appears when
  $0<B<\frac{E^2-1}{E^2+1}A$, and the critical surface is defined only 
  with a restricted value of $\theta$. This shape remind us  of
  the non-BPS 5-branes considered in Ref.\ \cite{Bary}. However, the
  large uniform electric field requires careful treatment (see
  Ref. \cite{Bachas}).}.

The study in \cite{CM} tells us that there must be another solution
which, together with the solution (\ref{sol!}), forms a single throat
connected smoothly. This would be an anti-brane with a hole. Now as
seen above, the shape of the critical surface is distorted, therefore
a naive replacement $(A,B)\rightarrow (-A,-B)$ does not lead to
another solution to be joined with the brane with the hole defined by
Eq.\ (\ref{cri}). Reluctantly, let us approximate the distance
$\Delta$ between the brane and the anti-brane by the doubled height of
the solution (\ref{sol!}) estimated at $\cos\theta=0$. Performing 
integration along the line defined by $\theta = \pi/2$, we have
\begin{eqnarray}
\frac{\Delta}2 \sim \frac1{\sqrt{1\!+\!E^2}}\cdot
\frac{(1\!+\!E^2)B - E^2A}{r_{\rm critical}(\theta\!=\!\pi/2)}\cdot
\int_1^\infty \!\!\frac{du}{\sqrt{u^4\!-\!1}}
\end{eqnarray}
The limit of large separation $\Delta\rightarrow\infty$ can be
attained by the following two different limits of the parameters $A$
and $B$: 
\begin{eqnarray}
{\rm (i)}&\quad   (B\!-\!A)  \rightarrow 0, \quad & (A\!+\!B)
\;\mbox{fixed}, \\ 
{\rm (ii)}&\quad  (B\!-\!A)\; \mbox{fixed},  \quad &
(A\!+\!B)\rightarrow \infty  .
\end{eqnarray}
where we suppose  $(A\!+\!B)^3, (A\!+\!B) \gg (B\!-\!A)$ when taking
these limits. In the limit (i), the critical radius of the solution
becomes narrow and approaches to the BPS configuration
(\ref{conf}). This solution would correspond to a fundamental string 
connecting the brane and the anti-brane. The limit (ii) is the
sphareron solution, discussed in detail in Refs.\ \cite{CM,death}. 

Although the general solution (\ref{sol!}) satisfies various
properties which will play a central role in evaluating the
annihilation as seen above, unfortunately this solution contains one
disappointing property: there is no configuration $X$ which yields the 
gradient (\ref{sol!})! In fact, one can chack that rotation of the
solution (\ref{sol!}) does not vanish: $\nabla\times(\nabla X) \neq 
0$. It follows that 
\begin{eqnarray}
\label{curl}
  \nabla\times(\nabla X) \; \propto \; E(B\!-\!A).
\end{eqnarray}
At a glance, the brane seems to have thickness.
At the spatial infinity, the rotation vanishes and the solution is
appropriately defined, however at finite $r$ we cannot say about the
location of the brane surface.  The meaning of the relation
(\ref{curl}) is investigated in the last section.



\section{Discussions}
\label{sec:dis}

In summary, we have investigated Born-Infeld dynamics in the uniform
electric field, for both BPS and non-BPS configurations. For the BPS
configuration, naive force balance ensures the stability of the
configuration, and the fluctuations around the configuration satisfy 
precisely the boundary conditions expected from the worldsheet
analysis \`{a} la Polchinski. For the non-BPS configuration, we have
solved explicitly and generally the equations of motion with
appropriate boundary conditions at the spatial infinity of the
brane. The obtained solution is the generalization of the throat
solution given in Refs.\ \cite{CM,Gibb}.

As mentioned in the last paragraph of Sec.\ \ref{sec:anni}, the
solution presented there, Eq.\ (\ref{sol!}), is not rotation-free.
This is actually a problem, since if we want to evaluate the brane -
anti-brane annihilation in terms of the Born-Infeld theory, the
path-integral should be performed over 
$X$ and not over $\nabla X$, hence the solution (\ref{sol!}) is not
relevant in the path-integral procedure. However, the solution
(\ref{sol!}) seems to be general, if we put the appropriate boundary
condition (\ref{bou})\footnote{The other solutions (\ref{anot2}) and
(\ref{solc}) which do not satisfy the boundary conditions are also
rotation non-free. Hence general non-BPS solutions $\nabla X$ with
$E\neq 0$ seem to exhibit the non-integrability.}.

This unsatisfactory result may be interpreted in the following way.
Eq.\ (\ref{curl}) indicates that the vanishing of the rotation of
$\nabla X$, which is a necessary condition for the integrability,
occurs when the solution is BPS ($A=B$), or, when there is no uniform
electric field ($E=0$). 
As for the BPS limit, it is known that the BPS solution of
the ordinary Maxwell-scalar system is actually the solution of fully
(higher derivative) corrected effective equations of motion of open
strings \cite{Th}. 
BPS configurations possess this sort of generality, hence exhibit fine 
properties even in the uniform electric field in the approximation
level of the Born-Infeld theory, as seen in Sec.\ \ref{sec:bc}.

Now, how about the existence of the uniform electric field? Let us
consider the case of a uniform magnetic field. As shown in Ref.\
\cite{Gibb}, if we turn on the magnetic field, the equations of motion 
(and constraint equations) are manifestly invariant under the
SO($2,1$) rotation in the plane spanned by $(\phi,\chi,X)$, where
$\chi$ is the magneto-static potential. Hence the solution (\ref{sol!})
is also the solution if $\E$ is replaced by the magnetic field $\B$
(with appropriately quantized magnetic charge $A$). For the uniform
magnetic field, it is 
known that the low energy effective action for the open strings is
{\it not} the ordinary super Yang-Mills, but the one with Moyal
brackets \cite{moyal}. Therefore, this non-commutativity of the Moyal
bracket may change the non-BPS solution to be integrable.

Finally, we comment on the relation to the supergravity
calculation. In Refs.\ \cite{LPT,Rey}, a test string in the background 
of supergravity solution of D3-branes are analyzed using the string
$\sigma$ model approach. It was found that fluctuations of the scalar
fields on the worldsheet feels the same form of the potential as in the
Born-Infeld analysis. In our case of the D3-brane with the uniform
electric field, this property would be able to be confirmed in a
similar manner. Recently, the supergravity solution representing 
the (F,D$p$) bound state was constructed \cite{Lu1,Lu2}. The
worldsheet action of the static probe string (which corresponds to the
tube part in Fig.\ \ref{fig:tube}) in this background is\footnote{We  
  turn on only the relevant two scalars, $X_3$ (parameterizing the 
  direction along the electric flux) and $X_\perp$ (the radial
  coordinate in the transverse directions).} 
\begin{eqnarray}
\label{lag}
  S=\int\! d\tau d\sigma
\left[
  (H')^{1/4}H^{-3/4}\sqrt{(X_3^{\prime})^2 +
    H(X_\perp')^2} -X_3'\frac{E}{\sqrt{1+E^2}}\frac1H
\right],
\end{eqnarray}
where $H$ and $H'$ are harmonic functions in the transverse space
\footnote{
The explicit forms of these functions are
\begin{eqnarray}
  H\equiv1+\frac{Q}{(X_\perp)^4}, \qquad
  H'\equiv1+\frac{Q}{(1+E^2)(X_\perp)^4}.
\end{eqnarray}
These functions give two distinct ``core''(this core is defined as the 
value of $X_\perp$ where the first term in the harmonic function (in
this case, 1) is comparable to the second term ($\sim 1/X_\perp^4$)). 
For this reason, the potential which the scalar feels, computed in the 
supergravity formulation, might be different from the one obtained in
Sec.\ \ref{sec:tra} and Sec.\ \ref{sec:long}. However, in the weak
coupling limit, two cores may overlap with each other and this potential
will reproduce the appropriate boundary conditions considered in this
paper.
}.
The second term in Eq.\ (\ref{lag}) stems from the
coupling of the probe string to the non-trivial NS-NS
2-from background produced by the fundamental strings condensed
in the D3-brane. Near the infinity of the transverse space, two
harmonic functions are approximated as $H\sim H' \sim 1$, hence the
static solution lead from the equations of motion is $X_3'=
kX_\perp'$, where $k$ is an integration constant.  Substituting this
relation into the action (\ref{lag}), we have 
\begin{eqnarray}
  S\sim
\int\!d\tau d\sigma X_\perp'
  \left(
    \sqrt{1+k^2}-k\frac{E}{\sqrt{1+E^2}}
  \right)
=
\int\!d\tau X_\perp\Bigm|_{\rm infinity}
\left(
    \sqrt{1+k^2}-k\frac{E}{\sqrt{1+E^2}}
  \right).
\end{eqnarray}
Noting that $X_\perp$ is positive, the action is minimized at the
value $k=E$. Thus at the spatial infinity we obtain the tilted
configuration of the string: $X_3'= EX_\perp'$, which is exactly
matched to the configuration considered in Sec.\ \ref{sec:stable}.

\vspace{.5cm}
\noindent
{\large\bf Acknowledgments}\\[.2cm]
I would like to thank K.\ Furuuchi, Y.\ Hyakutake, H.\ Kawai, T.\
Kugo, S.\ Nakamura and K.\ Yoshioka for valuable discussions and
comments.

\vspace{.5cm}
\appendix
\noindent
\section{Solution for fluctuation}

In this appendix, we present a way to solve the equation (\ref{delta}) 
using the expansion (\ref{exp}). Making an abbreviation
\begin{eqnarray}
{\cal O} \equiv \left( 1+E^2 + \frac{\pi^2 \g^2}{r^4} \right)\omega^2
+\frac{1}{r^2}\frac{\del}{\del r}\left(r^2\frac{\del}{\del r}\right),
\end{eqnarray}
the equation (\ref{delta}) is decomposed due to the orthogonality of
the Legendre functions as follows:
\begin{eqnarray}
&&  {\cal O}\eta_0 + \frac{2\epsilon}{r^2}\eta_1 =0,\nn\\
&&  \left({\cal O}-\frac{2}{r^2}\right)\eta_1 
      + \frac{2\epsilon}{r^2}\left(\eta_0+\frac25 \eta_2\right)
      =0,\label{dec}\\ 
&&  \left({\cal O}-\frac{6}{r^2}\right)\eta_2 
      + \frac{2\epsilon}{r^2}\left(\frac23\eta_1+\frac37
        \eta_3\right) =0, \quad\cdots\nn
\end{eqnarray}
where we put $\epsilon\equiv \pi E\g\omega^2$. Due to the structure of 
these equations, it is possible to expand $\eta_l$ further as 
\begin{eqnarray}
  \eta_l= \epsilon^l \eta_l^{(l)}+
  \epsilon^{l+2}\eta_l^{(l+2)}+\cdots,
\end{eqnarray}
where, in particular, $\eta_0^{(0)}$ is the zero mode of the operator
${\cal O}$. Then easily one can 
deduce that the leading terms of $\eta_l$ in each angular momentum $l$ 
satisfy relations 
\begin{eqnarray}
  \eta_1^{(1)}= \eta_0^{(0)},\quad
  \eta_2^{(2)}= \frac29 \eta_1^{(1)}, \quad\cdots,
\end{eqnarray}
therefore intrinsically all $\eta_l^{(l)}$ are identical with
$\eta_0^{(0)}$. The next-to-leading terms $\eta_l^{(l+2)}$ are
determined by 
evaluating the next-to-leading terms in the decomposed equations
(\ref{dec}). As an example, let us consider the first one:
\begin{eqnarray}
  {\cal O}\eta_0^{(2)}+\frac2{r^2}\eta_1^{(1)}=0.
\label{eta1}
\end{eqnarray}
At $r\sim 0$ ($y\sim\infty$), the region which represents the place
where the effect of the end point of the attached string is not
expected to appear, the operator ${\cal O}$ is approximated by 
\begin{eqnarray}
  {\cal O} \sim \frac{1}{\pi^2 \g^2 \omega^2}y^4
  \left(
    1+\frac{d^2}{dy^2}
  \right).
\end{eqnarray}
Therefore, using the asymptotic ($r\sim 0$) behavior of the solution 
$\eta_1^{(1)}=\eta_0^{(0)}\sim \exp (\pm iy)$ at the weak coupling
limit, we obtain a solution of (\ref{eta1}) as 
\begin{eqnarray}
  \eta_0^{(2)}\sim \frac1y\exp (\pm iy).
\end{eqnarray}
This does not contribute when we discuss the phase shift of the
boundary, since the magnitude of this mode is dumping fast enough at
$r\sim 0$. Owing to similar argument, the following relations 
can be derived:
\begin{eqnarray}
  \eta_l^{(l+2k)}\sim y^{-k}\exp (\pm iy) \quad \mbox{at $r\sim 0$}.
\end{eqnarray}
These modes ($k\geq 1$) are collectively represented in Eq.\
(\ref{sol}) as ``higher order terms''.

\newcommand{\J}[4]{{\sl #1} {\bf #2} (#3) #4}
\newcommand{\andJ}[3]{{\bf #1} (#2) #3}
\newcommand{\AP}{Ann.\ Phys.\ (N.Y.)}
\newcommand{\MPL}{Mod.\ Phys.\ Lett.}
\newcommand{\NP}{Nucl.\ Phys.}
\newcommand{\PL}{Phys.\ Lett.}
\newcommand{\PR}{Phys.\ Rev.}
\newcommand{\PRL}{Phys.\ Rev.\ Lett.}
\newcommand{\PTP}{Prog.\ Theor.\ Phys.}
\newcommand{\hep}[1]{{\tt hep-th/{#1}}}


\begin{thebibliography}{99}

\bibitem{Pol}
    J.\ Polchinski, 
    \J{\PRL}{75}{1995}{4724}, \hep{9510017}.

\bibitem{BI}
    M.\ Born and L.\ Infeld, 
    \J{\sl Proc.\ R.\ Soc.\ London,}{A144}{1934}{425}.

\bibitem{CLNY}
    C.\ G.\ Callan, C.\ Lovelace, C.\ R.\ Nappi and S.\ A.\ Yost, 
    \J{\NP}{B288}{1987}{525}.

\bibitem{BIeff}
    E.\ S.\ Fradkin and A.\ A.\ Tseytlin, \J{\PL}{B163}{1985}{123} ;\\
    E.\ Bergshoeff, E.\ Sezgin, C.\ N.\ Pope and P.\ K.\ Townsend,
    \J{\PL}{B188}{1987}{70}.

\bibitem{L}
    R.\ G.\ Leigh, 
    \J{\MPL}{A4}{1989}{2767}. 


\bibitem{CM} 
    C.\ G.\ Callan Jr.\ and J.\ M.\ Maldacena, 
    \J{\NP}{B513}{1998}{198}, \hep{9708147}.

\bibitem{Gibb}
    G.\ W.\ Gibbons, 
    \J{\NP}{B514}{1998}{603}, {\tt hep-th/9709027}.  

\bibitem{HHS}
    O.\ Bergman, \J{\NP}{B525}{1998}{104}, \hep{9712211} ; \\
    K.\ Hashimoto, H.\ Hata and N.\ Sasakura,
    \J{\PL}{B431}{1998}{303}, \hep{} \\ \hspace*{10mm}{\tt 9803127} ;
    \J{\NP}{B535}{1998}{83},  
    \hep{9804164} ; \\
    T.\ Kawano and K.\ Okuyama, \J{\PL}{B432}{1998}{338},
    \hep{9804139} ; \\
    K.\ Lee and P.\ Yi, \J{\PR}{D58}{1998}{066005}, \hep{9804174}.

\bibitem{death}
    K.\ G.\ Savvidy, 
    \hep{9810163}. 

\bibitem{death2}
    R.\ Emparan, 
    \J{\PL}{B423}{1998}{71}, \hep{9711106}.

\bibitem{add}
    D.\ Bak, J.\ Lee and H.\ Min, \J{\PR}{D59}{1999}{045011}, 
    \hep{9806149} ;\\
    S.\ Nojiri and S.\ D.\ Odintsov,
    \J{Int.\ J.\ Mod.\ Phys.}{A13}{1998}{2165}, 
    \J{ibid.}{A13}{1998}{4777} \hspace*{10mm}(erratum), \hep{9707142} ;
    \J{\PL}{B419}{1998}{107}, \hep{9710137} ;\\
    T.\ Kadoyoshi, S.\ Nojiri, S.\ D.\ Odintsov and A.\ Sugamoto,\\
    \hspace*{10mm}\J{Mod.\ Phys.\ Lett.}{A13}{1998}{1531},
    \hep{9710010}. 

\bibitem{Arf}
    H.\ Arfaei and M.\ M.\ Sheikh-Jabbari, 
    \J{\NP}{B526}{1998}{278}, \hep{9709054}.

\bibitem{Jab}
    M.\ M.\ Sheikh-Jabbari, 
    \J{\PL}{B425}{1998}{48}, \hep{9712199}.

\bibitem{Lu1}
    J.\ X.\ Lu and S.\ Roy,  
    \hep{9904112}.

\bibitem{Lu2}
    J.\ X.\ Lu and S.\ Roy,  
    \hep{9904129}. 


\bibitem{inter}
    A.\ A.\ Tseytlin, \J{Class.\ Quant.\ Grav.\ }{14}{1997}{2085}, 
    \hep{9702163} ;\\
    J.\ P.\ Gauntlett, G.\ W.\ Gibbons, G.\ Papadopoulos and P.\ K.\
    Townsend,\\ \hspace*{10mm}\J{\NP}{B500}{1997}{133}, \hep{9702202}
    ;\\  
    J.\ P.\ Gauntlett, \hep{9705011} ;\\
    N.\ Itzhaki, A.\ A.\ Tseytlin and S.\ Yankielowicz,
    \\ \hspace*{10mm}    \J{\PL}{B432}{1998}{298}, \hep{9803103} ;\\
    A.\ Hashimoto, \J{JHEP}{9901}{1999}{018}, \hep{9812159}.

\bibitem{DLP}
    J.\ Dai, R.\ G.\ Leigh and J.\ Polchinski,
    \J{\MPL}{A4}{1989}{2073}. 

\bibitem{susyBI}
    H.\ R.\ Christiansen, C.\ N\'{u}\~{n}ez and F.\ A.\ Schaposnik, 
    \J{\PL}{B441}{1998}{185}, \\ \hspace*{10mm}\hep{9807197}. 

\bibitem{Th} 
    L. Thorlacius, 
    \J{\PRL}{80}{1998}{1588}, \hep{9710181}. 

\bibitem{KYL}
    K.\ Lee, \J{\PL}{B445}{1999}{387}, \hep{9810110}.

\bibitem{Wit}
    E.\ Witten, 
    \J{\NP}{B460}{1996}{335}, \hep{9510135}.

\bibitem{SENNET}
    A.\ Sen, 
    \J{JHEP}{9803}{1998}{005}, {\tt hep-th/9711130}.

\bibitem{LPT} 
    S.\ Lee, A.\ Peet and L.\ Thorlacius, 
    \J{\NP}{B514}{1998}{161}, \hep{9710097}.

\bibitem{Rey}
    S.\ -J.\ Rey and J.\ Yee, 
    \hep{9803001}. 

\bibitem{savisavi}
    K.\ G.\ Savvidy and G.\ K.\ Savvidy, 
    \hep{9902023}.

\bibitem{tac}
    T.\ Banks and L.\ Susskind, \hep{9511194} ; \\
    A.\ Sen, \J{\sl JHEP}{9808}{1998}{012}, \hep{9805170}.

\bibitem{Hyaku}
    H.\ Awata, S.\ Hirano and Y.\ Hyakutake, 
    \hep{9902158}.

\bibitem{Bary}
    C.\ G.\ Callan, A.\ Guijosa and K.\ G.\ Savvidy, 
    \hep{9810092} ;\\
    C.\ G.\ Callan, A.\ Guijosa, K.\ G.\ Savvidy and O.\ Tafjord, 
    \hep{9902197}.

\bibitem{Bachas}
    C.\ Bachas and M.\ Porrati, 
    \J{\PL}{B296}{1992}{77}, \hep{9209032}.  

\bibitem{moyal}
    M.\ R.\ Douglas and C.\ Hull, \J{JHEP}{9802}{1998}{008},
    \hep{9711165} ;\\
    C.\ Hofman and E.\ Verlinde, 
    \J{JHEP}{9812}{1998}{010}, \hep{9810116} ;\\
    M.\ M.\ Sheikh-Jabbari, \J{\PL}{B450}{1999}{119}, \hep{9810179} 
    ;\\
    C.\ -S.\ Chu and P.\ -M.\ Ho, \hep{9812219}.

\end{thebibliography}
\end{document}